\documentclass[a4paper]{llncs}

\usepackage[english]{babel}
\usepackage[utf8x]{inputenc}
\usepackage[T1]{fontenc}
\usepackage{xcolor}

\usepackage[a4paper,top=3cm,bottom=2cm,left=3cm,right=3cm,marginparwidth=1.75cm]{geometry}

\usepackage{graphicx}

\usepackage{array}
\newcolumntype{L}[1]{>{\raggedright\let\newline\\\centering\arraybackslash\hspace{0pt}}m{#1}}

\title{An Open Source Solution for \\ Smart Contract-based Parking}
\author{
Nikolay Buldakov \inst{1}, Timur Khalilev \inst{1}, Salvatore Distefano \inst{2} and Manuel Mazzara \inst{1}\\
}

\institute{
Innopolis University, Russian Federation\\
\email{\{n.buldakov, t.khalilev, m.mazzara\}@innopolis.ru}
\and University of Messina \\
\email{sdistefano@unime.it}
}

\begin{document}
\maketitle

\authorrunning{Nikolay Buldakov, Timur Khalilev et al.}

 \abstract{
This paper discusses an open source solution to smart-parking in highly urbanized areas. Interviews have been conducted with domain experts, user stories defined and a system architecture has been proposed with a case study. Our solution allows independent owners of parking space to be integrated into one unified system, that facilitates the parking situation in a smart city. The utilization of such a system raises the issues of trust and transparency among several actors of the parking process. In order to tackle those, we propose a smart contract-based solution, that brings in trust by encapsulating sensitive relations and processes into transparent and distributed smart contracts.
}

\section{Introduction}
\label{sec:intro}

Due to the rapidly increasing urbanization, the management of many processes in cities has become an extremely cumbersome issue. With the current constantly growing urbanisation rate, which is about to reach 60\% \cite{mcgranahan2014urbanisation}, the traditional ad-hoc approaches to handling processes in cities do not provide sufficient results anymore.  One of such processes is traffic. The increasing population leads to increasing traffic congestion. 

There are several approaches to tackling this problem. For example, in 1970s many countries decided to widen the roads and create new ones. In Hague some canals were drained and covered to for this purpose \cite{hagueCanals}. However, with time empirical and statistical studies have shown that increasing the number of lanes favors the increasing traffic congestion. A more modern approach is to analyse various factors that lead affect congestion and operate with them. This has led to smart management of streetlights \cite{lau2015traffic}, introduction of payed parking lots and to the decrease in the number of lanes. 
    
In addition to that, studies have shown that search for a unoccupied parking stall also drastically affects this issue as cars spend a great matter of time circling around in attempts to park \cite{giuffre2012novel}. Thus, smart parking management can bring a significant improvement of the traffic situation to many cities. 

There are several potential areas for improvement in the current parking situation.  First of all, in every district there are pieces of unused land, either in private or in public possession. Providing easy means of renting out such property for parking can simplify the process of creating new parking lots. Moreover, collection and aggregation of information about the current occupancy of all parking lots can significantly simplify the process of finding a free stall, thus decreasing traffic congestion. Another important issue is the trustworthiness of a potential parking system. With the current solutions on the market, such as \cite{yang2012smart} sensitive information is stored in a centralised way, which makes it vulnerable and simplifies malicious intrusion. Finally, various parking providers need to be aggregated under one universal system, that will guarantee unbiased treatment of all parking lots. This will allow potential seekers of parking lots to observe the complete situation in the city instead of seeing only partial information. This work aims at creating an open source solution in order to tackle the aforementioned problems.

\subsection*{Problem Statement}

There is a city. As any city it has a governing authority. This can be a City Hall, a municipality or it can be divided up into districts with local councils. It can even be in a form of smart governance as described in \cite{meijer2016governing}. This authority will further be referred to as an Administrator. The Administrator is responsible for authorising processes in the city. There are also people and companies, who posses pieces of land suitable for parking. They will be referred to as Landlords. A landlord can be a juridical person or a natural person, it does not matter for the system. A landlord want to provide their land for parking in order to profit from parking fees. Besides, there are entities called Tenants. A tenant is a company or a person who will to whom a landlord rents out a subset of their parking land based on some conditions.

\begin{figure}[tb]
\centering
\includegraphics[scale=0.15]{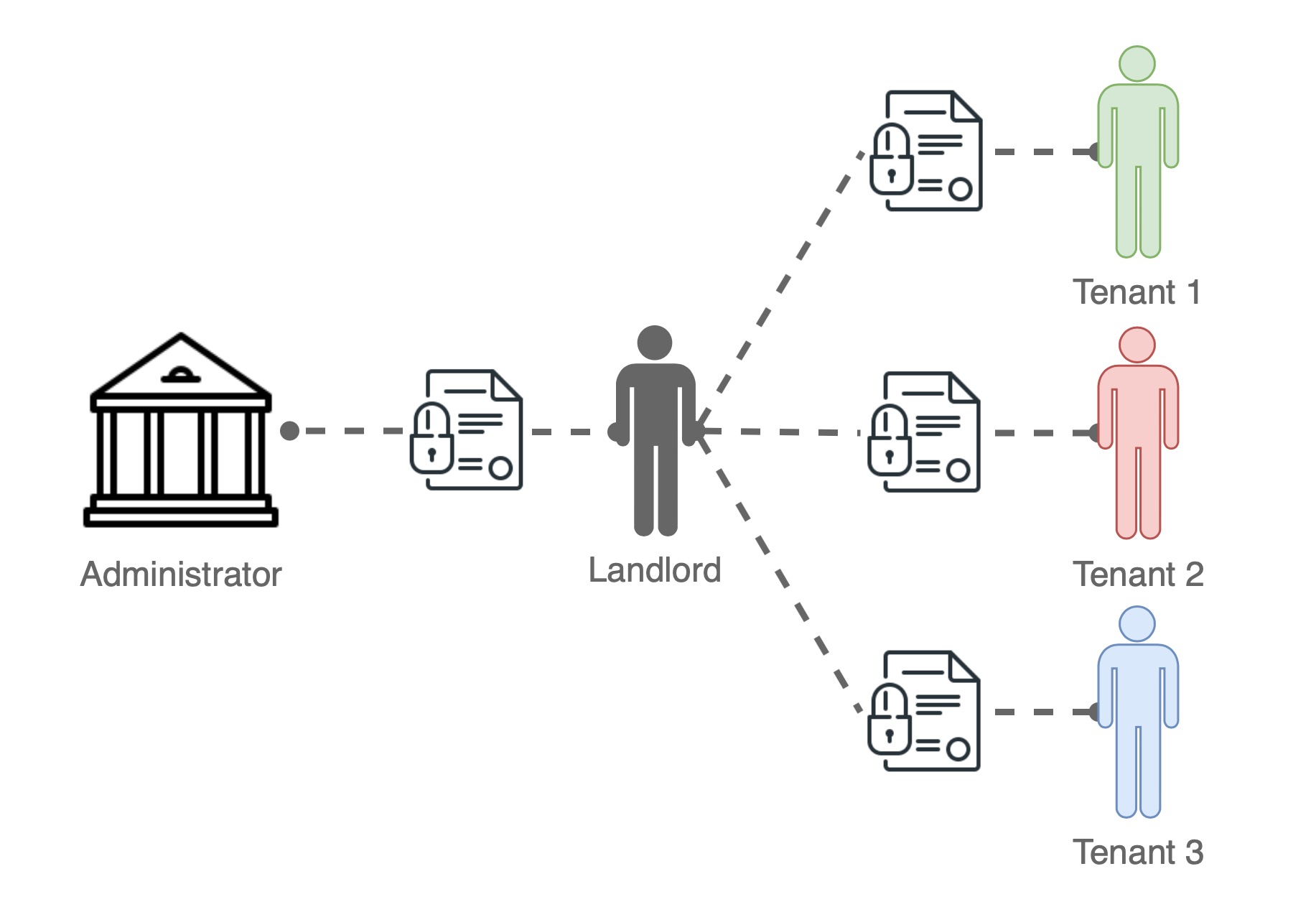}
\caption{Contracts' hierarchy}
\label{fig:problemStatementContracts}
\end{figure}

Now that all the roles are defined, one can observe the relations connection different roles in figure \ref{fig:problemStatementContracts}. In order to provide their land for parking, a landlord must first sign a contract with the administrator (there is always only one for each area in the city). The process starting from a request to create such contract (from the landlord's side) up until its signing by both parties will be called \textit{landlord registration}. This process involves checking whether the landlord is eligible to renting out their property, as well as negotiating on the details of the contract. Eventually, the signed contract contains such details as the tax which the landlord guarantees to pay, information about the registered land, duration of the contract and so on. The contract can be further amended, given that the both parties are satisfied with the amendment.
\label{sec:proposedSolution}

The landlord, in turn, can not only provide their land for parking, but also they can rent it out to tenants. In this case, the income from parking will not go directly to the landlord, however they receive guaranteed rent payments and, possibly, a percentage from each parking transaction. Similarly to the case of administrator - landlord relations, the agreement between each tenant and the landlord is preserved in for of a contract, where all the details of rent are described, such as the duration of the contract, the rent fee, the frequency of payments, the landlord's share in each parking transaction and, most importantly, the exact borders of the area rented to each tenant. Again, all these details can be amended later, given the consensus of both parties. 

\subsection*{Assumptions about parking lot}

\begin{figure}[tb]
\centering
\includegraphics[scale=0.15]{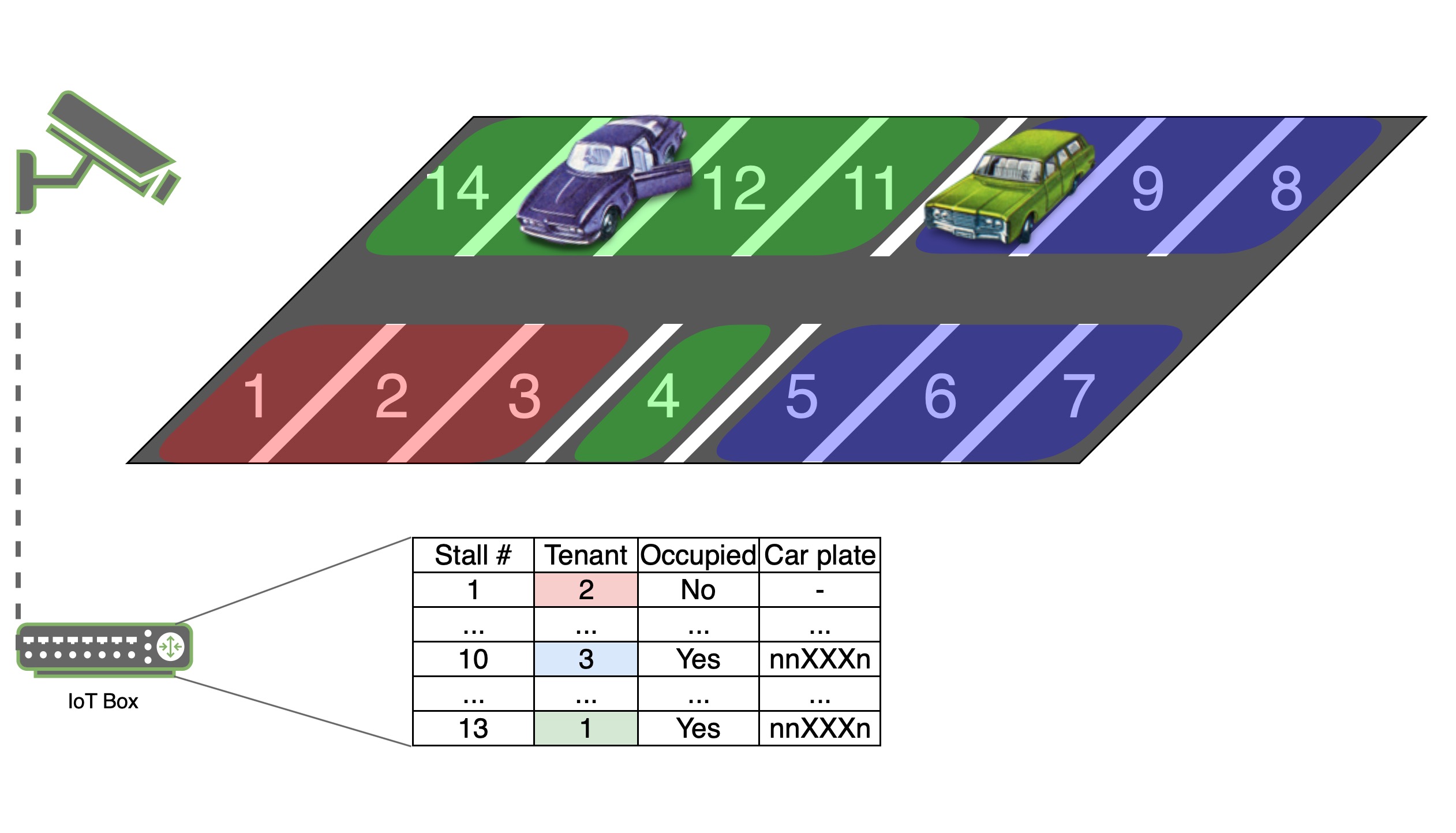}
\caption{Example of the assumed parking lot}
\label{fig:problemStatementParking}
\end{figure}

There is a set of preconditions, which this study assumes parking lots to hold. First of all, the parking areas have to be divided up into a set of spots. These spots will be further referred to as \textit{parking stalls}. This partition can be represented by road marking. Moreover, there has to be a tracking device which will make the parking lot fully observable. A particular example can be a camera with an IoT-powered box, produced by Park Smart \cite{parkSmart} as described in their patent \cite{smart2014allocating}. This device has to run image processing algorithms in spot in order to tell at each moment of time, which parking stalls are occupied (a particular implementation is provided in \cite{acharya2018real}) and what the plate numbers of the parked cars are. A visualisation of these assumptions can be found in Figure \ref{fig:problemStatementParking}

A real example could be a mall with a big parking zone. Stalls can be partitioned among the businesses that are present in the mall. Big companies, like Ikea could rent the closest stalls in order to ease transportation of goods from the store to cars. Restaurants could want to offer reduced fares to their visitors for parking and so on. There can also be stalls that are not rented by any company, in this case a car parked there pays according to the pricing policy of the landlord, i.e. the mall.

\subsection*{Proposed solution}
\label{sec:proposedSolution}

The proposed solution presented in this paper aims to tackle the problems described above by using blockchain smart contracts. It assumes a set of roles to be present is the city. One of them is a city hall, that will be responsible for the deployment of the system as well as setting several policies in accordance with the local legislation. Another role is a landlord who possesses a piece of land. A landlord can be a juridical person or a natural person, it does not matter for the system as long as their land satisfies the constraints describer in the Assumptions section. This is a common situation when a landlord does not want to maintain the land fully themselves and rents its parts out (link), in this case the proposed system allows a landlord to partition their land as they desire and set complex paying policies, that will ensure the agreed upon conditions at the time of payment for parking. The last role is a driver interested in parking their car. They can observe free spots available at every parking lot and the pricing policy and choose the most suitable one. Payment is also covered conducted through the system, using a micro-payment channel, that makes it possible to enforce and guarantee the distribution of money among the landlords, the city hall and the tenants.
\par

\section{Related Work}
\label{sec:related}

By the time of writing, various smart parking systems have been proposed in the literature, that focus on various aspects of parking.  In principle, they could be divided into two groups: centralized solutions and decentralised solutions.

\subsection*{Centralized solutions}
There is a garden variety of solutions present both in the market and research, which attempt to prove a framework to manage smart parking. However, the major drawback of most of them is their centralised nature. Many of them rely on a particular supreme entity, their "source of truth".  In \cite{khanna2016iot} an parking system is proposed that relies on IoT devices and exchanges data by the means of TCP/IP protocol. However, the application is deployed on only one server, that makes it vulnerable to issues with security, trustability and single point of failure. 

\subsection*{Decentralized solutions}
\begin{figure}[tb]
\centering
\includegraphics[width=\textwidth]{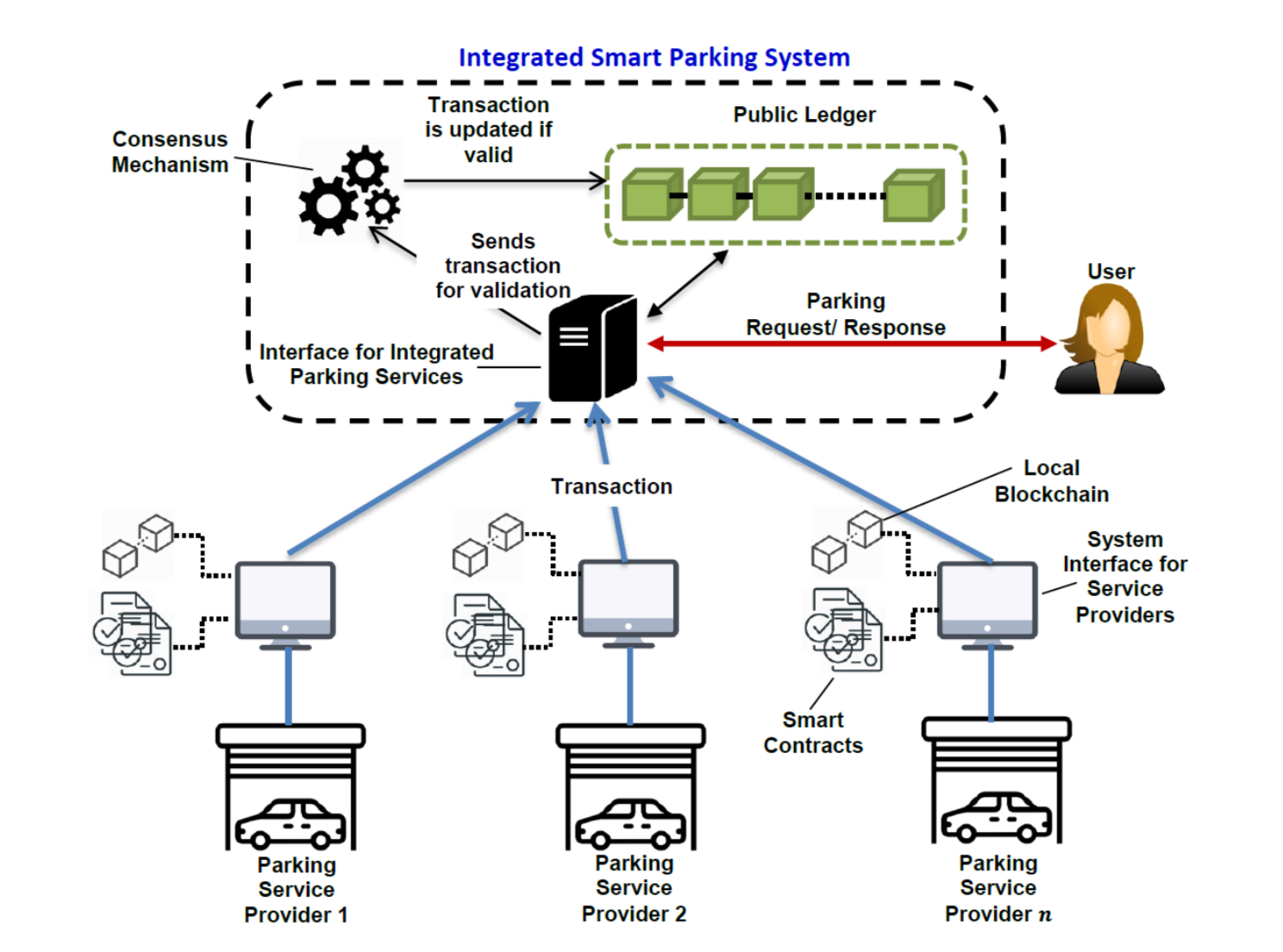}
\caption{Overview of Integrated Smart Parking System, taken from \cite{ahmed2019blockchain}}
\label{fig:OverviewofIntegratedSmartParkingSystem}
\end{figure}

Other works propose architectures that utilise the blockchain solutions as their backbone. All centralised solutions share a common set of problems that make them undesirable for a practical use. The most important of such issues is trustworthiness. All participants of such systems have to rely on a particular entity, that will posses the power of accessing, chaining and removing their data, with no guarantee at the time of signing an agreement or registering  that the owner of the system will obey to the promised rules. Lack of transparency is another significant shortcoming. When the implementation is not exposed to the end users, they cannot validate that a particular solution is guaranteed to function in accordance with their needs and that it does not have any back-doors. Finally, considering the ongoing competition on any market, including parking, a centralised solution can be unfair to some participants if the owner is biased towards some companies.
In \cite{ahmed2019blockchain} a system is proposed for incorporating blockchain into regular parking scenarios. This paper presents a layered architecture that standardizes the typical architecture of integrated smart parking system. It also describes the major components of the system as well as provides an overview of possible work-flows of the solution. In this system there are three major participants: parking service provider, blockchain network, and user. The network enjoys the standard blockchain characteristics: a consensus algorithm, a shared public ledger, running a local copy of the ledger at each parking area. The overview schema of the system can be seen in \ref{fig:OverviewofIntegratedSmartParkingSystem}. The architecture in the described solution consists of four layers: application layer, network layer, transaction layer, and physical layer. This layout is layout of technologies can be considered the current state of art and is adapted in the solution, proposed in this thesis work. Finally, the authors of the paper describe two main workflows supported by their system: 1. Search and Request of a parking spot and 2. Parking Provider Operations in the system. Even though the paper describes a useful basic concept it supports a very limited number of scenarios and thus cannot be considered as realistic. 

\par Another blockchain-powered system is presented in \cite{hassan2019parkchain}. It aims to take into account the process of setting up the parking space for the land-owner. For this they a landlord has to go through government validated process and procedures such as Land Registry and Government mandate external system. The external control is conducted by presenting another entity, named Oracle. In addition to that, they tackle the problem of leasing lands to the so-called contractors by distributing non-fungible tokens among them for a particular period of time. These tokens are issued to the certified landlords by the government. The important advantage of this solution is that it describes in great details peculiarities related to how governing authorities and lease of land can be integrated into a blockchain system. However, many processes, which actually being novelty to this work, are supposed to happen off-chain, which is a serious limitation for the true decentralisation of the system.

\cite{hu2019parking} presents registration, search, rent, payment modules built by the means of the BCOS smart contracts framework. The main novelty of this system is the use of one-time pad mechanism and a group-signature mechanism together with a union of privileged nodes in order to conceal the details of each transactions and thus preserve users' privacy while the privileged nodes can still access users' data. One-time padding mean that instead of receiving a real address for each transaction, users, instead, receive a new generated address each time. This way, the users' addresses are not exposed. A group signature mechanism allows signing a transaction on behalf of a group, which hides the mapping between a transactions and users. Other concepts implemented in this system are fairly similar to the aforementioned works, so the privacy mechanisms are the main beneficial feature of the paper. 

Finally \cite{ferreira2018blockchain} introduces a gamification approach, which simplifies the process of finding a parking lot by encouraging users to take part in the reporting process, by using the data history in order to calculate which places are more likely to be empty and by providing a frame work that does not assume or require any particular installation of an infrastructure. In this system the concept of mobile crowdsensing plays a pivotal role. In order to ensure the involvement of drivers, the system assigns special points to each driver for their collaboration (such as installation of a special beacon or reporting on the number of free parking spaces in a street). Pedestrians can also participate by reporting through a mobile app. These points can be exchanged for free parking minutes. The authors claim their approach to bring a variety of benefits including a faster search for parking (leading to lower CO2 emissions), easy deployment of the solution as it does not require much an existing infrastructure. Although their reasoning is valid, there is almost no information in the paper how the blockchain technology was incorporated.

\subsection*{Discussion}
Although the works mentioned in this literature review present a variety of parking solutions, a major limitation is in the fact that all of them lack flexibility for the end users. There is a variety of possible configurations of the system present, for example with a governing agency involved \cite{hassan2019parkchain} or without \cite{ahmed2019blockchain} but none of them allows the community to choose what type of configuration they need. The same issue arises with the pricing policy. All papers consider some pricing policy present and even fluctuating depending on some factors as time and demand \cite{ahmed2019blockchain}, however the details are omitted of how exactly the pricing is organized as well as how shares are distributed among different entities in the system. Finally, the topic of leasing or renting out the land is not thoroughly defined, in particular, how exactly it can be done through the system. In \cite{hassan2019parkchain} a token system is proposed but the details of how exactly the partitioning of a land will be tackled in the system are not presented.

\section{System Design}
\label{sec:design}

The objective of this work is to take into account drawbacks of the current centralised parking solutions, as well as limitations of the decentralised parking systems and to design and build a system that would address those issues. The system will utilise smart contracts and take into consideration some realistic requirements gathered by interviewing people from the industry. 

\subsection*{Interviewing Industry representatives}
Any proper system-development commences with gathering important requirements of the system. And as \cite{requirementsRechniques} states
"one might use questionnaires or interviews to explore what problems members see as most important, how members place themselves in various classification schemes, etc."  Thus, an interview has been conducted with a representative of Park Smart \cite{parkSmart}. Park Smart is a smart parking solution company founded in Italy, Messina in 2014 and its main aim is to implement a software-based technology able to analyse, in real time, the availability of space in in-street parking. As the company has already been in the field for a few years, it was valuable to question them about their experience and to learn what are the main requirements for a parking system.

The interview has shown some important aspects of modern smart parking system and it has also revealed some common shortcomings of the existing system. For instance, if a company wants to use this solution for paid parking, they can provide their end point to Park Smart in order to receive money though the system. This approach has inspired the idea of service providers in our system. Moreover, the interviewee has confirmed our concern about centralisation of current systems. In fact, Park Smart itself has a patent about blockchain architecture for their system, however, it has not been implemented yet. Besides, the importance of flexible pricing policies has been highlighted. Companies have different rates at different time of the day, therefore this part of functionally is to be flexible. Finally, we found out what particular information about users and processes is needed for such a system and this has affected our decisions of what to store in the proposed solution

\subsection*{Product backlog}

In order to comprehend the needs of various user-roles in the system, user stories have been gathered. The roles that we considered are the following: \textit{Administrator}, \textit{Landlord}, \textit{Tenant}, \textit{Driver}, and \textit{Service Provider}. 

The user stories have revealed a particular set of features, that every role needed. Here we extract the key functionality from the user stories. First of all, both driver and parking providers (landlords and tenants) require a safe trustworthy way of conducting payments in such a way that both parties are guaranteed that the amount of funds transferred is equivalent to the provided service. Besides, landlords and tenants want to ensure that the renting contracts they make among them are respected by both parties and that such contracts can be extended or amended over time. Parking providers also want the system to postulate equal treatment for all parking lots, so that there are no privileged ones, promoted be that system. Finally, the administrative authority requires to have privileged over the system (bounded by conditions of contracts made with other parties).

In total the backlog is constituted by more than 30 user stories, which were considered during the design and implementation phases.

\subsection*{System Design Overview}

There are several participants in the system: \textit{landlord}, \textit{tenant}, \textit{driver or car}, \textit{IoT box} installed in each parking system, \textit{administrator} and as a governing authority. The parking systems consists of thin clients (such as a mobile or a web app) and each participant can access their part of functionality through the thin client. Every thin client is connected to a local EVM (Etherium Virtual Machine) node, which in turn is connected to the rest of the Etherium network. The local node has an instance of ledger, shared by all nodes, and also hosts smart contracts.

\begin{figure}[tb]
\centering
\includegraphics[scale=0.1]{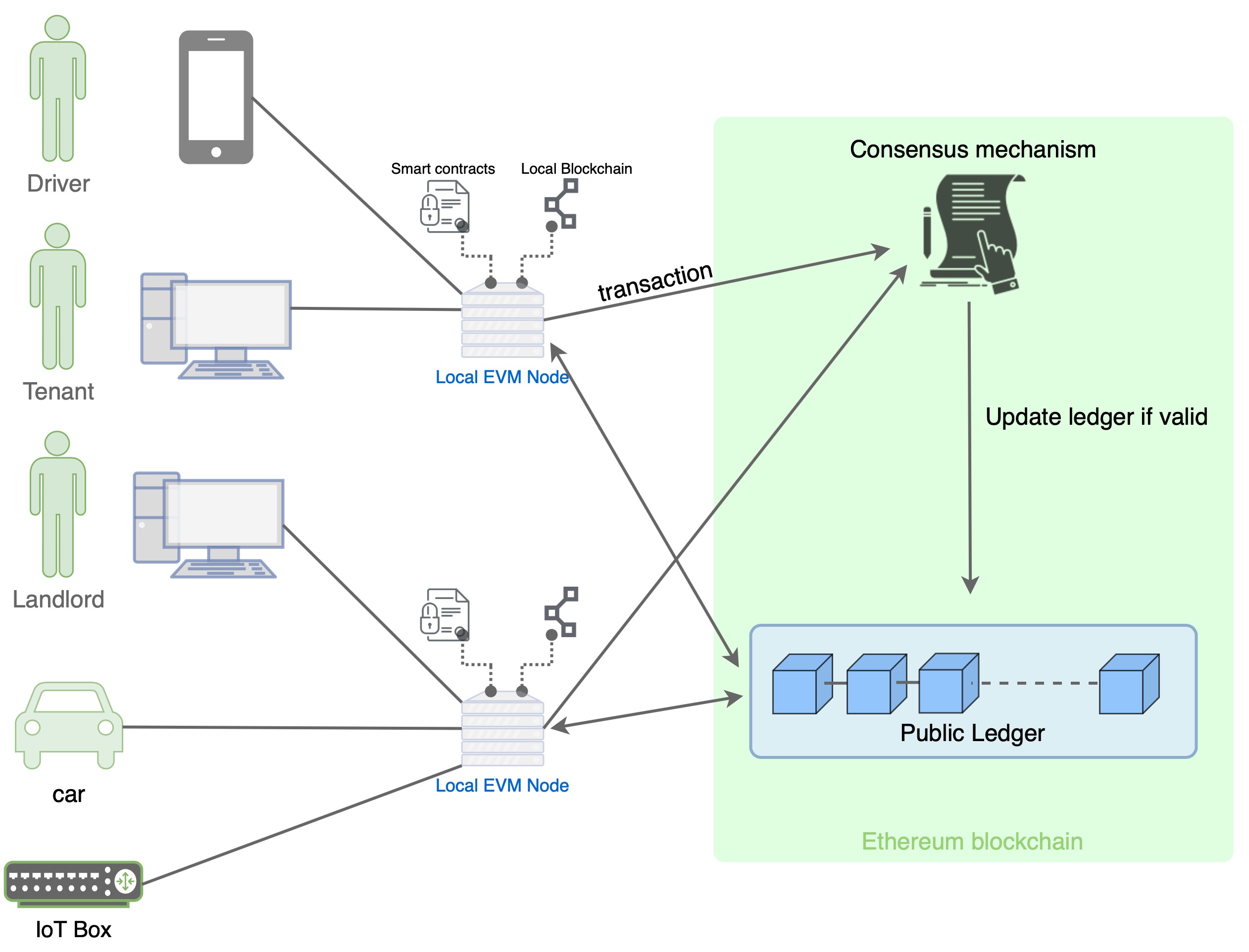}
\caption{System Overview}
\label{fig:overview}
\end{figure}

Whenever a participant want to perform an action (e.g. initiate parking or update number of free parking sports), the think client tell the EVM to invoke a corresponding method of some smart contract. This action is performed in the form of a transaction on the blockchain. The blockchain network contains a public ledger and updates the public ledger with the valid transactions only. A consensus mechanism is used to verify the transactions. Once the new transaction is verified, the public ledger is updated and the local copies of the ledger in each node are synchronised and the new state of the parking system can be observed by all participants. This overview is visualised in \ref{fig:overview}

\subsection*{Smart contracts}
Smart contracts are the back-end part of the system. The logic is distributed among several contracts, which correspond to logical components of the system. Each contract encapsulates its sensitive data and controls access to its functions. This way only authorised entities can perform actions in the system, even though the contracts are publicly exposed. Some of the contracts are deployed by actors of the system, others are deployed automatically in existing contracts. The decision whether a contract is deployed from the outside of the system or within contracts depends on two questions:

\begin{enumerate}
    \item Is the nature of the contract static or dynamic?
    \item Should the system validate and control the deployment process of a contract?
\end{enumerate}

The nature of a contract matters because some contracts are deployed once and remain active in the system for a long time, whereas others are needed only for a short amount of time and after their expiration, they cannot be used inside the system anymore. An example of a static contract could be the Parking System contract, which is deployed by the authorities and the life cycle of this contract defines the life cycle of the system itself. On the other hand, an example of a contract of a more dynamic nature could be a Payment Channel. Whenever a new parking process commences, the system automatically generates a payment channel. Such contract has a short lifetime and cannot be reused, thus the burden of its creation lies upon the shoulders of a static contract – Parking Provider.

The latter question tackles the problem of validation. Contracts can be inherited and their behaviour changed by other parties. For example, the Parking Lot contract can be modified to surpass the tax obligation. That is why, one requests the system to deploy such a contract. The system verifies all information related to the contract and only after that deploys it. For other contracts, on the contrary, extensions are allowed. For instance, Payment Policy defines a set of methods to be implemented to calculate a price of a particular parking, however, it is up to a parking provider how exactly the price will be calculated. Thus, a parking provider is entitled to extend and deploy their own payment policy and supply the address of this policy to the system.

\paragraph{Parking System}
\paragraph{}
This is the first contract to be deployed to launch the system. It is deployed by an administrative authority (e.g. a city hall) and it implements the main managing functionality of the system. The parking system contract is responsible for registration of new parking lots and cars as well as for storing addresses of all other entities in the system. Whenever some request cannot be fulfilled automatically by the contract, it saves the request and pends administration's approval. Moreover, this contact provides means of control and management of the whole system. Thus, the system cannot exists without this contract.

\paragraph{Parking Provider}
\paragraph{}
Parking provider is an abstract contract that cannot be used on its own but should be extended in order to be used. It implements the logic of the parking process and stores information about what parking stalls it possesses and what cars are parked at the stalls. In other words, whenever a car wants to park, the code of this contract is used. In order to park a car sends funds to this contract along with a stall number and till what time it wants to be parked. The contract calculates the price, checks if the funds sent are sufficient and whether a parking stall is free and after that creates a payment channel. Optionally parking provider can also have a set a service providers, in this case a car can specify which service provider it is using and a fraction of the payments will go to the service provider.

\paragraph{Car}
\paragraph{}
This contract preserves general information about a car, such as its plate number, account address of its owner and the rating of the car in the system. In addition to that, it stores information about parking, such as whether the car is parked now or not, the address of the payment channel, parking history, etc. If one needs some information about a car, this is the contract to address.

\paragraph{Parking Lot}
\paragraph{}
Parking lot extends Parking Provider contract and is deployed by the parking system contracts at a request of a landlord and with an approval of the administrator. This contract adds functionality to partition the parking lot among several tenants. In order to do so as tenant creates a request to this contract. If the landlord approves the request, the contract creates a renting contract between the landlord and the tenant. It also contains general information about the parking lot, such as its rating, address and so on.

\paragraph{Tenant}
\paragraph{}
Tenant also extends Parking Provider contract. Unlike Parking Lot contract, this contract does not need permission from the system to be deployed as it interacts only with a parking lot. Tenant can set its own payment policy and it also stores the address of its contract with a parking lot.
\paragraph{Service Provider}
\paragraph{}
Conceptually a service provider is a third-party entity that collaborates with a parking provider in order to place information about the parking provider's stall on their platform. Thus, the contract only facilitates the interaction between a service provider and a parking provider. It contains information about a service provider and about parking provider it collaborates with.
\paragraph{Renting Contract}
\paragraph{}

Renting contract implements a relationship between a parking lot and a tenant in such a way that both parties are guaranteed that their agreement will be respected in the system. The contract is deployed when the parking lot contract approves registration of a new tenant. After that the tenant reviews the contract and if conditions are satisfactory, the renting process begins. The contract also contains logic for penalties in case of a delayed payment and supports the introduction of amendments from both parties. In case of mutual agreement, such amendments can modify the behaviour of the contract.

\paragraph{Payment Channel}
\paragraph{}
Payment channel implements logic for micro payments between two parties, in such a way that both parties are guaranteed to receive the promised funds. One party opens the contract and sends funds to it. After that the funds are locked in the contract and neither of the parties can access them. Micropayments are conducted off chain, therefore they are not part of this contract. Once they payment process is completed, the receiving party sends the last encrypted message to the contract and the contract uses encryption algorithms to verify that the party is entitled to receive the funds. If that is the case, funds are sent to the receiver's address. The detailed description of the principles behind this contract can be found in \ref{sec:microPaymentChannel}  

\paragraph{Payment Policy}
\paragraph{}
Payment policy is a protocol that defines two methods: get payment rate at a particular time and get a total price for parking given the start and the end time. It is up to parking providers to define their own parking policies, depending on their business goals. A particular implementation of this protocol is provided. It implements a policy that can have a specific rate depending on an hour and a day of the week.

\subsection*{Parking system front-end}
The front-end part of the parking system is a distributed application in form of web, mobile or embedded software installed on mobile phones and computers of the Administrator, landlords, tenants and drivers as well as in IoT boxes at parkingLots In future they will be referred to as think clients. Every thin client provides a user interface or an API for interaction with the system. User interfaces are used by landlords, administrator, tenants and drivers. The API is used by the IoT box or a smart car. The front-end part is responsible for translating users' actions into requests to the smart contracts. 

\subsection*{Communication between the smart contracts and the parking system}

\begin{figure}[tb]
\centering
\includegraphics[width=\textwidth]{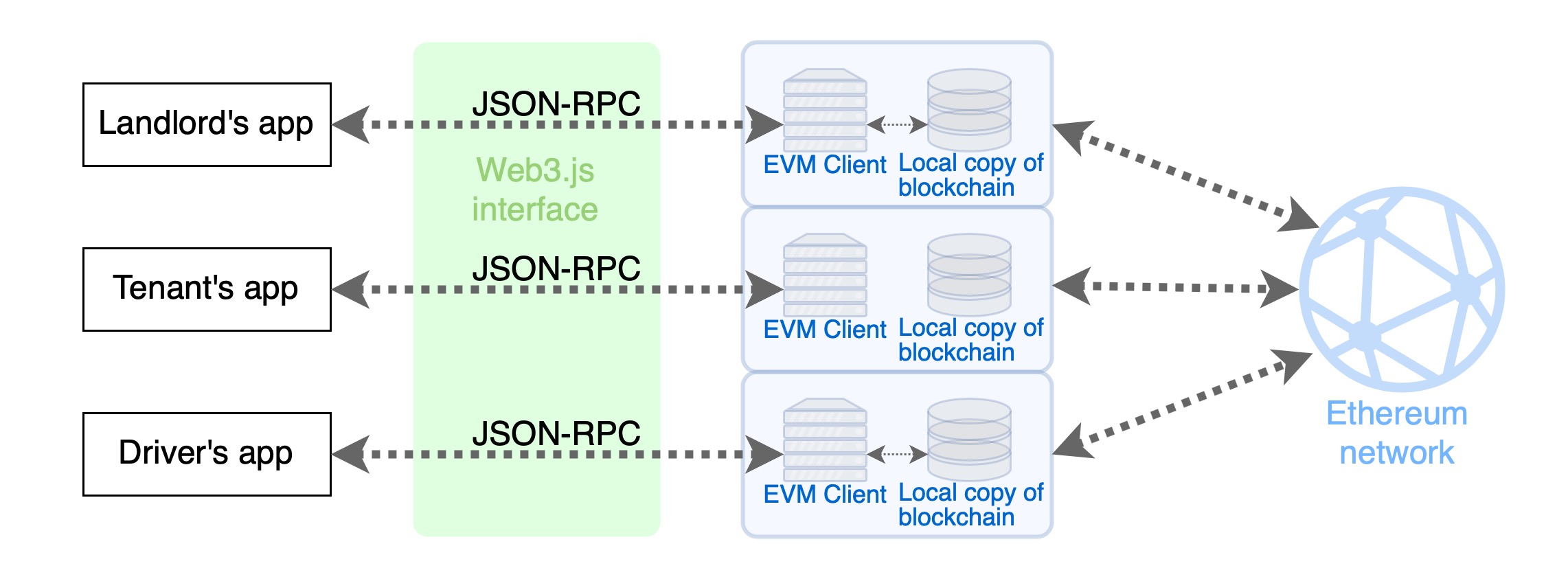}
\caption{Communication with the blockchain}
\label{fig:web3Interaction}
\end{figure}

In order for the front-end distributed application to communicate with the EVM, there has to be as specific data-interchange format. This format is called JSON-RPC \cite{json2013json}. It is utilised through a JavaScript library called Web3.js \footnote{https://github.com/ethereum/web3.js/}, which allows you to interact with a local or remote Ethereum node using a HTTP or IPC connection \cite{web3}. As depicted in Figure \ref{fig:web3Interaction} Every front-end module of the distributed system communicates mostly with the their own smart contract and this contract, in turn, is responsible for the interaction with the Parking System module and other modules. 

\subsection*{Micro-payment channel}
\label{sec:microPaymentChannel}
Whenever car gets parked at a parking lot, there is a mutual agreement between the parking lot and the car: the car receives a service and transfers money for parking. In a traditional system, the payment is conducted in advance for a particular period of time,  \cite{litman2018parking}. Such system lacks trustworthiness, as a car has to blindly trust the parking provider, that the service will be provided fully and will not be interrupted earlier, the parking provider, in turn, has to ensure that the car is not parked for longer than the payed period, this requires additional control that brings an additional cost. Finally, parking is not a discreet process unlike the pricing policy in traditional parking systems, which results in drivers often times paying more than they should.

This problem can be solved by implementing a micro-payment channel by the means of a smart contract and direct signed messages. The high-level picture of such payment includes three steps: 
\begin{enumerate}
    \item Driver deploys a contract with locked funds of some amount that exceeds any possible parking period.
    \item Driver emits messages directly to the service provider. The messages are the "micro-transactions" containing how amount is transferred. The parking lot can check validity of the messages as they are signed by car.
    \item After the parking process is over, the service provider can use the last message to retrieve the claimed amount of funds from the contract. The remained of the locked funds are sent back to the car.
\end{enumerate}

\begin{figure}[]
\centering
\includegraphics[scale=0.1]{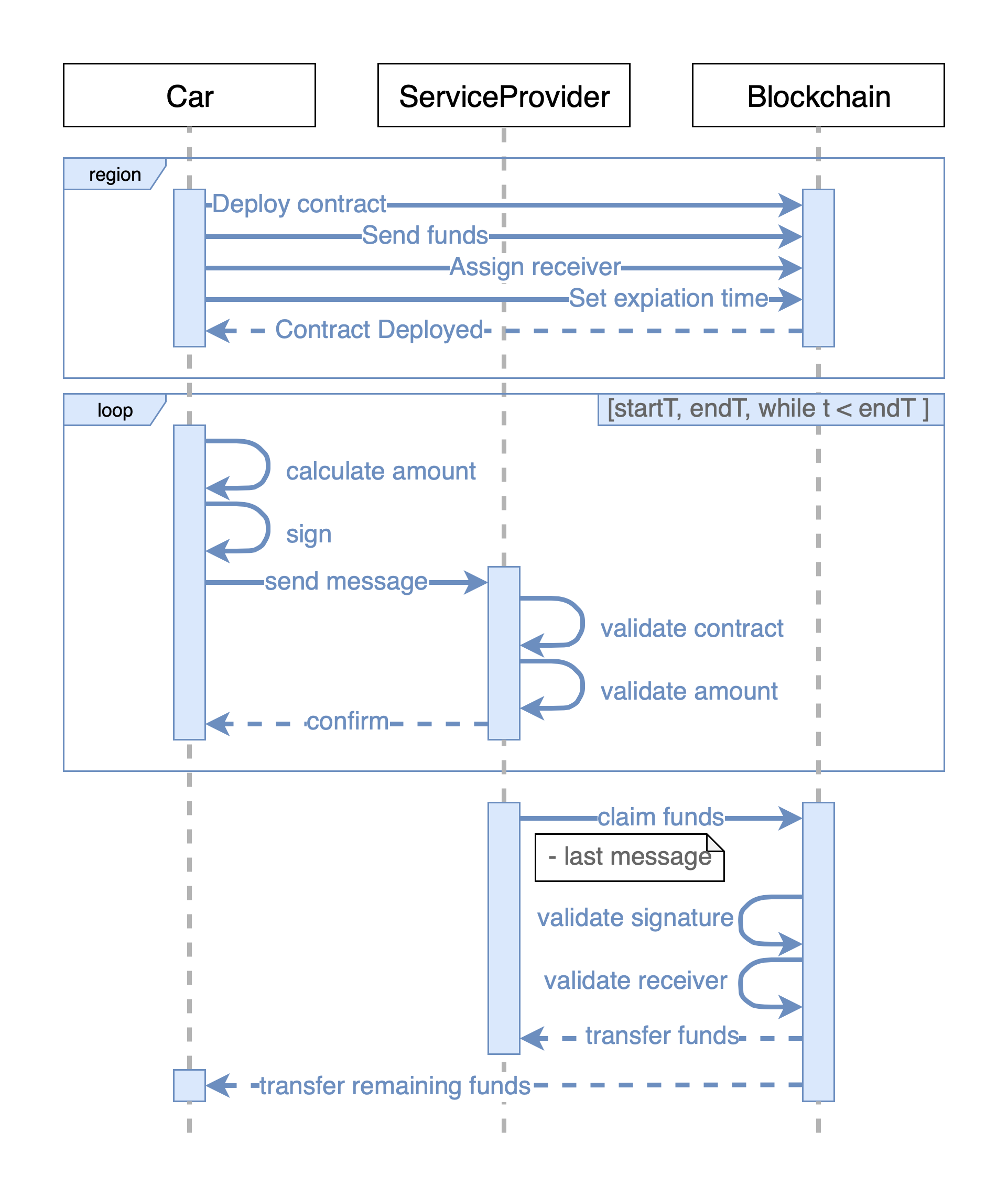}
\caption{Sequence diagram of micro-payments over blockchain}
\label{fig:seqMicropayment}
\end{figure}

It is important to mention that in this scenario only steps 1 and 3 happen on the blockchain and the most intense process of micro-payments is conducted off-chain through a peer-to-peer communication \cite{peerjs}. Thus, the parties do not have to conduct costly blockchain transactions (gas in Etherium terminology \cite{dannen2017introducing}) for each payment. Only two transactions are required: to deploy the contract (by car) and to claim the end of the payment period (by the parking lot). Moreover, the transaction fee will be paid by the service provider, thus removing any additional fees from the car. This process is visualised in \ref{fig:seqMicropayment}

\section{Conclusions}
\label{sec:conclusions}

A significant amount of research has been conducted so for on smart parking. In this paper we have examined some of the recent work related to this topic and propose a solution. Interviews have been conducted with domain experts, user stories defined and a system architecture has been proposed with a case study. Our solution allows independent owners of parking space to be integrated into one unified system, that facilitates the parking situation in a smart city. The utilization of such a system raises the issues of trust and transparency among several actors of the parking process. In order to tackle those, we propose a smart contract-based solution, that brings in trust by encapsulating sensitive relations and processes into transparent and distributed smart contracts. From the architecture point of view services and, in particular, microservices \cite{Dragoni2017,MazzaraBDR20,BDDL2020} and their composition, orchestration \cite{Mazzara:phd,Mazzara2005}, reconfiguration \cite{MazzaraADB11} and recovery \cite{Mazzara2009} have not been discussed. All these are open issues that need to be investigated in future.

\bibliographystyle{plain}
\bibliography{MS}  
\end{document}